# Classification of Kantowski-Sachs and Bianchi Type III space-times according to their proper conformal vector fields

Ghulam Shabbir, Suhail Khan and M. Ramzan

Faculty of Engineering Sciences,

GIK Institute of Engineering Sciences and Technology,

Topi, Swabi, NWFP, Pakistan.

Email: shabbir@giki.edu.pk

**Abstract**

We investigate proper conformal vector fields in non conformally flat Kantowski-Sachs and Bianchi type III space-times using direct integration technique. Using the above mentioned technique we show that very special classes of the above space-times admit proper conformal vector fields.

## 1. INTRODUCTION

Due to the significant role of symmetry in the Einstein theory of general relativity, different symmetries have been investigated by many authors [1,2 and reference there in]. Unlike Killing, homothetic and affine symmetries, conformal symmetry is difficult to study because it lacks the linearity property. These difficulties are discussed in [3-6]. Some more general results on the Lie algebra and dimensions of conformal vector fields are given in [7]. The aim of this paper is to find the existance of conformal vector fields in the non conformally flat Kantowski-Sachs and Bianchi type III space-times. The conformal vector field which preserves the metric structure upto a conformal factor carries significant information and interest in Einstein's theory of general relativity. It is therefore important to study this symmetry. In this paper a direct integration technique is used to study conformal vector fields in the non conformally flat Kantowski-Sachs and Bianchi type III space-times. Throughout $M$ represents a four dimensional, connected, hausdorff space-time manifold with Lorentz metric $g$ of signature (-, +, +, +). The curvature tensor associated with $g_{ab}$, through the Levi-Civita connection, is denoted in component form by $R^a{}_{bcd}$ and the Weyl tensor components are $C^a{}_{bcd}$. The usual covariant, partial and Lie derivatives are denoted by a semicolon, a comma and the symbol $L$, respectively. Round and square brackets denote the usual symmetrization and skew-symmetrization, respectively. The space-time $M$ will be assumed non conformally flat in the sense that the Weyl tensor does not vanish over any non empty open subset of $M$.

Any vector field $X$ on $M$ can be decomposed as

$$X_{a;b} = \frac{1}{2} h_{ab} + F_{ab} \qquad (1)$$

where $h_{ab}(=h_{ba}) = L_X g_{ab}$ and $F_{ab}(=-F_{ba})$ are symmetric and skew symmetric tensors on $M$, respectively. Such a vector field $X$ is called conformal vector field if the local diffeomorphisms $\eta_t$ (for appropriate $t$) associated with $X$ preserve the metric structure up to a conformal factor i.e. $\eta_t^* g = \psi\, g$, where $\psi$ is a nowhere zero positive function on some open subset of $M$ and $\eta_t^*$ is a pullback map on some open subset of $M$ [2,8]. This is equivalent to the condition that

$$h_{ab} = \psi\, g_{ab},$$

equivalently,

$$g_{ab,c} X^c + g_{cb} X^c{}_{,a} + g_{ac} X^c{}_{,b} = \psi\, g_{ab}, \qquad (2)$$

where $\psi : U \to R$ is the smooth conformal function on some subset of $M$, then $X$ is a called conformal vector field. If $\psi$ is constant on $M$, $X$ is homothetic (proper homothetic if $\psi \neq 0$) while if $\psi = 0$ it is Killing. If the vector field $X$ is not homothetic then it is called proper conformal.

## 2. Main Results

Consider the space-times in the usual coordinate system $(t,r,\theta,\phi)$ with line element [1]

$$ds^2 = -dt^2 + A(t)dr^2 + B(t)\left[d\theta^2 + f^2(\theta)d\phi^2\right], \qquad (3)$$

where $A$ and $B$ are no where zero functions of $t$ only. For $f(\theta) = \sin\theta$ or $f(\theta) = \sinh\theta$ the above space-time (3) become Kantowski-Sachs or Bianchi type III space-times, respectively. The above space-time admits four independent Killing fields which are [9]

$$\frac{\partial}{\partial r},\ \frac{\partial}{\partial \phi},\ \cos\phi\frac{\partial}{\partial \theta} - \frac{f'}{f}\sin\phi\frac{\partial}{\partial \phi},\ \sin\phi\frac{\partial}{\partial \theta} + \frac{f'}{f}\cos\phi\frac{\partial}{\partial \phi}, \qquad (4)$$

where prime denotes the derivative with respect to $\theta$. A vector field $X$ is said to be a conformal vector field if it satisfies equation (2). One can write (2) explicitly using (3)

$$X^0{}_{,0} = \psi(t,r,\theta,\phi), \qquad (5)$$

$$X^0{}_{,1} - A X^1{}_{,0} = 0 \qquad (6)$$

$$X^0{}_{,2} - B X^2{}_{,0} = 0 \qquad (7)$$

$$X^0{}_{,3} - B f^2(\theta) X^3{}_{,0} = 0 \qquad (8)$$

$$\frac{\dot{A}}{2A} X^0 + X^1{}_{,1} = \psi(t,r,\theta,\phi) \qquad (9)$$

$$A X^1{}_{,2} + B X^2{}_{,1} = 0 \qquad (10)$$

$$A X^1{}_{,3} + B f^2(\theta) X^3{}_{,1} = 0 \qquad (11)$$

$$\frac{\dot{B}}{2B} X^0 + X^2{}_{,2} = \psi(t,r,\theta,\phi) \qquad (12)$$

$$X^2{}_{,3} + f^2(\theta) X^3{}_{,2} = 0 \qquad (13)$$

$$\frac{\dot{B}}{2B} X^0 + \frac{f'(\theta)}{f(\theta)} X^2 + X^3{}_{,3} = \psi(t,r,\theta,\phi), \qquad (14)$$

where dot denotes the derivative with respect to $t$. Equations (5), (6), (7) and (8) give

$$X^0 = \int \psi(t,r,\theta,\phi)\,dt + C^1(r,\theta,\phi),$$

$$X^1 = \int \frac{1}{A}\left(\int \psi_r(t,r,\theta,\phi)\,dt\right)dt + C_r^1(r,\theta,\phi)\int \frac{1}{A}\,dt + C^2(r,\theta,\phi),$$

$$X^2 = \int \frac{1}{B}\left(\int \psi_\theta(t,r,\theta,\phi)\,dt\right)dt + C_\theta^1(r,\theta,\phi)\int \frac{1}{B}\,dt + C^3(r,\theta,\phi),$$

$$X^3 = \int \frac{1}{Bf^2(\theta)}\left(\int \psi_\phi(t,r,\theta,\phi)\,dt\right)dt + C_\phi^1(r,\theta,\phi)\int \frac{1}{Bf^2(\theta)}\,dt + C^4(r,\theta,\phi),$$

(15)

where $C^1(r,\theta,\phi), C^2(r,\theta,\phi), C^3(r,\theta,\phi)$ and $C^4(r,\theta,\phi)$ are functions of integration. In order to determine $C^1(r,\theta,\phi), C^2(r,\theta,\phi), C^3(r,\theta,\phi)$ and $C^4(r,\theta,\phi)$ we need to integrate the remaining six equations. To avoid details, here we will present only results, when the above space-times (3) admit conformal vector fields. It follows after some tedious and lengthy calculations that the following possibilities exist when the above space-times (3) admit conformal vector fields which are:

**Case (1)**

Here the space-time (3) becomes

$$ds^2 = -dt^2 + V^2(t)\left[e^{-N(t)}dr^2 + d\theta^2 + f^2(\theta)d\phi^2\right],\qquad (16)$$

where $N(t) = \int \frac{2c_1}{V(t)}\,dt,\ V(t) = \int \psi(t)\,dt + c_2$ and $c_1, c_2 \in R(c_1 \neq 0)$. The conformal vector fields in this case are

$$X^0 = V(t),\ X^1 = rc_3 + c_7,\ X^2 = c_4\cos\phi + c_5\sin\phi,$$
$$X^3 = -c_4\frac{f'}{f}\sin\phi + c_5\frac{f'}{f}\cos\phi + c_6,\qquad (17)$$

where $c_3, c_4, c_5, c_6, c_7 \in R$. One can write the above equation (17) after subtracting Killing vector fields as

$$X = (V(t), rc_3, 0, 0).\qquad (18)$$

The above space-time (16) admits five independent conformal vector fields which are given in equation (17) in which one is proper conformal see equation (18) and four are independent Killing vector fields. The conformal factor in this case is

$$\psi(t) = \frac{dV(t)}{dt}.$$

**Case (2)**

In this case the space-time (3) becomes

$$ds^2 = -dt^2 + V^2(t)\left[dr^2 + d\theta^2 + f^2(\theta)d\phi^2\right], \qquad (19)$$

where $V(t) = \int \psi(t)dt + c_2$ and $c_2 \in R$. The conformal vector fields in this case are

$$X^0 = V(t), \quad X^1 = c_7, \quad X^2 = c_4 \cos\phi + c_5 \sin\phi,$$
$$X^3 = -c_4 \frac{f'}{f}\sin\phi + c_5 \frac{f'}{f}\cos\phi + c_6, \qquad (20)$$

where $c_4, c_5, c_6, c_7 \in R$. One can write the above equation (20) after subtracting Killing vector fields

$$X = (V(t),0,0,0). \qquad (21)$$

Here, the above space-time (19) admits five independent conformal vector fields which are given in equation (20) in which one is proper conformal see equation (21) and four are independent Killing vector fields. The conformal factor in this case is $\psi(t) = \dfrac{dV(t)}{dt}$.

## References

[1] H. Stephani, D. Kramer, M. A. H. MacCallum, C. Hoenselears and E. Herlt, Exact Solutions of Einstein's Field Equations, Cambridge University Press, 2003.
[2] G. S. Hall, Symmetries and Curvature Structure in General Relativity, World Scientific, 2004.
[3] R. F. Bilyalov, Sov. Physics, **8** (1964) 878.
[4] L. Defrise-Carter, Commun. Math. Physics, **40** (1975) 273.
[5] G. S. Hall, J. Math. Physics, **31** (1990) 1198.
[6] Michael Steller, Ann. Glob. Annal Geom, **29** (2006) 293.
[7] G. S. Hall and J. D. Steele, J. Math. Physics, **32** (1991) 1847.
[8] R. M. Wald, General Relativity, The University of Chicago Press, 1984.
[9] G. Shabbir and A. B. Mehmood, Modern Physics Letters A, **22** (2007) 807.